# Alignment and preliminary outcomes of an ELT-size instrument to a very large telescope: LINC-NIRVANA at LBT

M. Bergomi[*][a,b], L. Marafatto[a,b], V. Viotto[a,b], C. Arcidiacono[c,b], J. Farinato[a,b], K.K. R. Santhakumari[d], R. McGurk[e], M. Dima[a,b], H. Baumeister[d], T. Bertram[d], J. Berwein[d], P. Bizenberger[d], F. Briegel[d], F.Kittman[d], A. Conrad[f], R. Ragazzoni[a,b], T.M. Herbst[d]

[a] INAF - Osservatorio Astronomico di Padova, Vicolo dell'Osservatorio 5, 35122 Padova, Italy
[b] ADONI - National Laboratory for Adaptive Optics, Italy
[c] INAF-Osservatorio Astronomico di Bologna, Via Ranzani 1, 40127 Bologna, Italy
[d] Max-Planck-Institut für Astronomie (MPIA), Königstuhl 17, 69117 Heidelberg, Germany
[e] Carnegie Observatories, 813 Santa Barbara St, Pasadena, CA 91101, USA
[f] LBTO, University of Arizona, 933 N. Cherry Ave, Room 552, Tucson, AZ 85721, USA

## ABSTRACT

LINC-NIRVANA (LN) is a high resolution, near infrared imager that uses a multiple field-of-view, layer-oriented, multi-conjugate AO system, consisting of four multi-pyramid wavefront sensors (two for each arm of the Large Binocular Telescope, each conjugated to a different altitude). The system employs up to 40 star probes, looking at up to 20 natural guide stars simultaneously.

Its final goal is to perform Fizeau interferometric imaging, thereby achieving ELT-like spatial resolution (22.8 m baseline resolution). For this reason, LN is also equipped with a fringe tracker, a beam combiner and a NIR science camera, for a total of more than 250 optical components and an overall size of approximately 6x4x4.5 meters.

This paper describes the tradeoffs evaluated in order to achieve the alignment of the system to the telescope. We note that LN is comparable in size to planned ELT instrumentation. The impact of such alignment strategies will be compared and the selected procedure, where the LBT telescope is, in fact, aligned to the instrument, will be described. Furthermore, results coming from early night-time commissioning of the system will be presented.

**Keywords:** LINC-NIRVANA, First-light, Alignment, LBT, Pyramid WFS, GLAO, MCAO, Commissioning

## 1. INTRODUCTION

LINC-NIRVANA (LN)[1][2] is a high resolution, near infrared imager, which was installed at the Large Binocular Telescope (LBT) in September 2016. LN takes advantage of a multiple field of view, layer-oriented, multi-conjugate AO system [3] consisting of four pyramid[4] WaveFront Sensors (WFSs - two for each arm of the LBT, each conjugated at a different height). Each Ground-layer Wavefront Sensor (GWS) searches for up to 12 NGSs within a 2.8'-6' annular Field of View (FoV, in order to allow a 2' unvignetted FoV to the high-layer sensor). The sensor drives the 672-actuator LBT adaptive secondary mirror (M2)[5] at a maximum frequency of 1 KHz, to correct atmospheric turbulence, conjugated about 100 meters above the telescope pupil. Each High-layer Wavefront Sensor (HWS) searches for up to 8 NGSs within a 2'FoV and drives the 349-actuator commercial Xinetics Deformable Mirror, conjugated about 7.1 km above the telescope pupil. Each sensor optically co-adds the light coming from the NGSs to allow the use of fainter guide stars and consequently increase the sky coverage. While the footprints at the ground-layer are completely overlapped, this is not true for the HWS, as described in [6]. LN (see [7] for more details, key elements and optical path) has an overall size of approximately 6x4x4.5 meters and a weight of more than 10 tons, making it comparable in dimensions to future ELTs instruments.

---

[*] maria.bergomi@oapd.inaf.it

The complex adaptive optics system and the tight requirements make the alignment of LN [8] extremely challenging, and imply a full planning of the alignment procedures and strategies from subsystems up to alignment to the telescope, which will be discussed in the next section.

## 2. ALIGNMENT TO THE TELESCOPE

Given the challenges presented by the internal alignment strategy of such a complex instrument, at the time of the first full internal alignment of the system, discussions and options were evaluated and the first decision was taken: the telescope would have to be aligned to the instrument rather than the instrument to the telescope. This obviously also affected the system internal alignment procedure [9]. In fact, the two arms not only had to match the internal alignment requirements but also they considered the requirement coming from telescope mirrors adjustment capabilities coupled to the uncertainty on the final LN position after its installation on the telescope, which given the size and weight of the overall instrument could only be roughly corrected on-site.

Some preliminary hints of the strategy of alignment of the LBT mirrors to the instrument were given by the LN Pathfinder experiment [12], although the annular field of view of the GWS imposed a further challenge, since the sensor requires an off-axis light source.

After reaching the LBT site, the subsystems alignment was internally verified, all subsystems were installed onto the LN bench and the optical path leading to scientific camera and HWSs aligned, as described in [8]. At this stage, the GWSs were co-aligned to the rest of the LN optical path, through adjustments of the annular mirrors' tilt and position. To perform this activity, we used a telescope simulator, an F/15 source (nicknamed Magic Lantern, see Figure 1) able to span about 90% of the whole circular 6' FoV.

In such a way, during LBT alignment to LN, we could overcome the GWS annular FoV challenge (absence of on-axis surface) and exploit the availability for the HWS's path of an on-axis field of view.

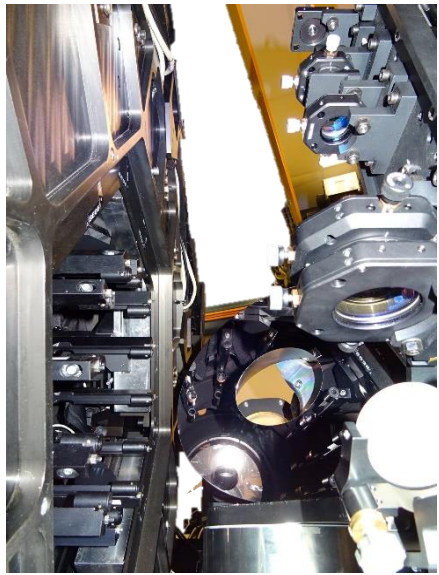

Figure 1: F/15 light source (aka Magic Lantern, visible on the right of the picture), used to tune the annular mirror (visible in the lower-mid part of the picture) in order to align the GWSs sensors (visible in the left of the picture) to the rest of the LN bench.

The internally aligned and characterized instrument was installed at the Gregorian Focal Station of LBT in September 2016 (Figure 2).

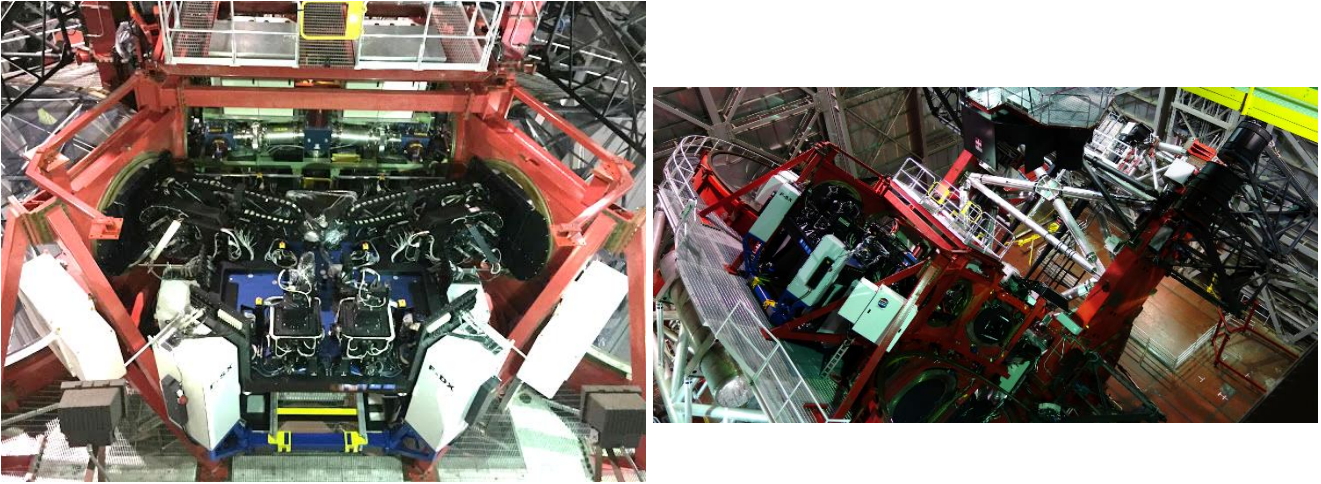

Figure 2: LINC-NIRVANA installed at the bent Gregorian focal station of LBT

Discussions on the alignment strategy of the telescope to LN lead to prefer the safest alignment strategy, to be fine tuned during the on-sky commissioning nights. The identified procedure exploited existing instruments/systems already aligned to the telescope and of the camera looking at the inner 2' FoV of LN, more specifically the Patrol Camera and the CCD39 of the HWSs.

We took advantage of the presence of:

- the ARGOS [10] calibration unit central reference fiber (focal plane, Figure 3 left panel)
- the LUCI [11] Auto-Guiding and Wavefront sensing (AGW) unit, already aligned to the telescope, with a reference position "hot-pixel" defined
- the 4 LEDs located around M2 (pupil-plane). Their image as appears on the CCD50 is shown in Figure 3, middle panel
- the HWS Patrol camera and CCD39 (WFS camera)

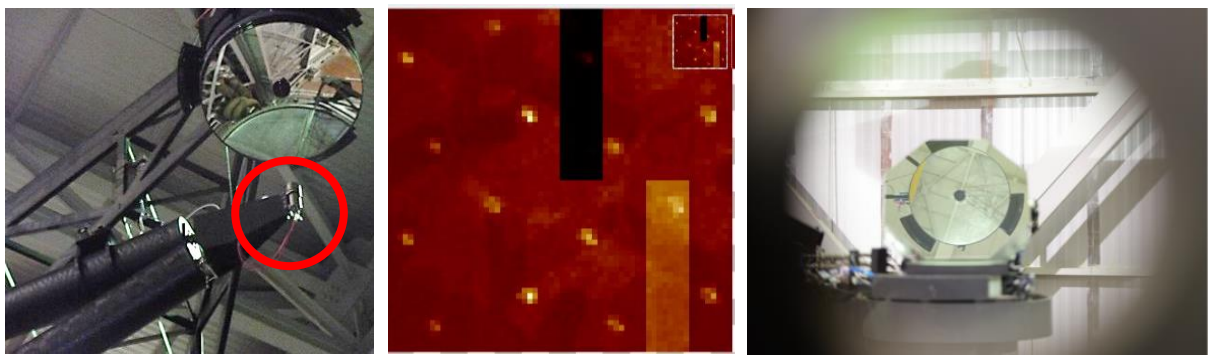

Figure 3: *Left*: the ARGOS calibration unit (circled in red) pointing toward LBT M2. *Middle*: LEDs located around M2 imaged around the 4 pupils location on the CCD50 of the GWS. *Right*: M2 and M3 mirrors as seen from behind the Annular mirror.

During the procedure, the LBT M3 mirror is set either to illuminate LUCI focal stations in night-time configuration (LUCI mode) or to illuminate LINC-NIRVANA ones (LN mode). In LN mode, the correct conic constant needs to be applied to M2. The ARGOS arm is swung into the optical path before the alignment and kept in place with the central reference fiber source switched on. Coma-free pointing implies rotations of the two mirrors as a single rigid body, maintaining a collimated model.

The main steps for the alignment procedure (for each telescope side) are:

- in LUCI mode, coma is minimized and the decentering of M2 is adjusted until the centroid of the ARGOS central reference source spot falls on the LUCI AGW hot-pixel;
- in LN mode, the M2 hexapod is adjusted in order to focus the ARGOS fiber on the LN Patrol Camera and coma-free pointing movements are used to adjust ARGOS position on the HWS Patrol Camera reference pixel;
- in LN mode, the 4 LEDs are switched on, the HWSs probes are inserted in the field and, looking at CCD39, the 4 LEDs barycenter are centered with respect to the pupil tilting M3. An alternative would be to insert a HWS probe at the on-axis position and center the light, in order to equally illuminate the 4 pupils;
- the two LN mode steps were iterated until they converged;
- at the end of the procedure, in LUCI mode, it was double-checked that the ARGOS source was still on the hot-pixel.

The obtained parameters for the M2 and M3 mirrors were stored in a configuration file to be used later during daytime and night-time activities.

The alignment of the SX side was completed in October 2016, while the DX side was set back by a temporary failure of the CCD39. A rough alignment could be achieved through the use of a paper screen, but a more robust alignment was performed in March 2017.

## 3. DAY-TIME AND NIGHT-TIME COMMISSIONING ACTIVITIES

One of the key activities performed during daytime calibrations concerned the Interaction Matrix (IM) calibration for the LN WFSs. Strategies and lessons-learned from LN Pathfinder experiment [12] were applied. In particular, we note that in order to track off-axis stars during night observations, the wavefront sensors of LINC-NIRVANA, are equipped with a derotator. Due to the fact that the ASM does not rotate, there is a continuously changing, rotating relationship between the sensor sub-apertures and the ASM actuators. As for Pathfinder, only a few IMs at different rotation angles were computed, through the mapping between known shapes (Karhunen-Loève modal basis), applied to the ASM with a push-pull scheme, and the signal measured by the Pyramid wavefront sensor. An interpolated matrix is produced and then rotated numerically to obtain synthetic reconstructors for every degree of rotation. During LN commissioning, the setup for the GWS IMs calibration on each telescope side was an on-axis source, obtained using a fiber (fed by white light) installed at the telescope prime-focus location (taking advantage of the LBTO retro-reflector support) shining light toward the ASM, reflected to the LBT tertiary mirror and into LINC-NIRVANA. More details on the GWSs IMs calibration can be found in [13]. At the same time, the HWS calibrations[15] took place on the other side of the instrument, through an internal fiber plate source. Some of these tests occurred also through remote connections.

Even though the Pathfinder experience helped us in some ways, the many reflections, different rotation angles and foci, made challenging to understand the sky orientation of the LN several focal planes. For this reason, a visualizer for different focal planes orientation was realized[14] and night-time was used to confirm the relationship between focal plane and sky, our plate-scale and pointing at stars. The software for the automatic pointing, acquisition and tracking of the guide stars was also improved during the on-sky activities, along with a fine tuning of the algorithm to determine commissioning fields with sufficient bright stars and the proper asterism distribution in the field.

## 4. ON-SKY PRELIMINARY OUTCOMES

The first photons of a 1$^{st}$ magnitude star reached the scientific and technical detectors on November 22$^{nd}$ 2016. We do not formally consider this as a first light, as it only involved some subsystems, but it did confirm the success of our alignment strategy. However, some further mirror alignment tuning was involved, looking at a 5$^{th}$ magnitude star, to avoid saturation of the detectors. Two snapshots taken after SX telescope side collimation and before DX side collimation on the HWS Patrol cameras and scientific camera are shown in Figure 4.

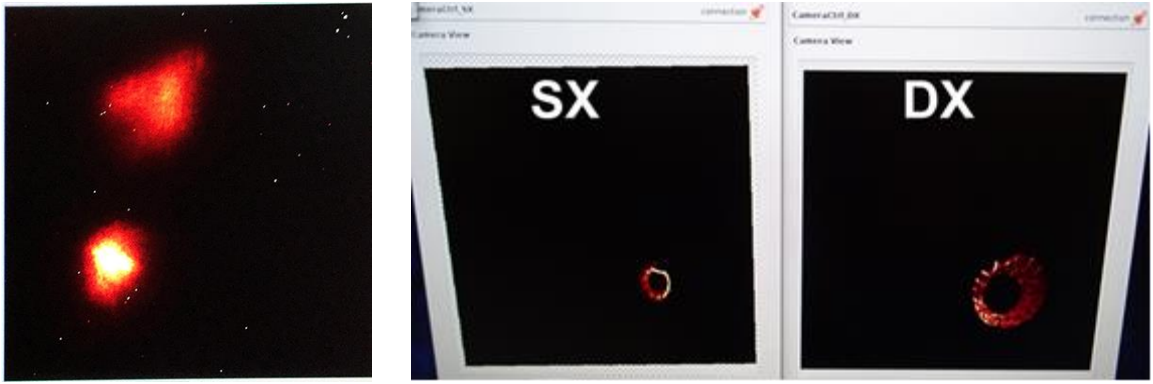

Figure 4: *Left*: first photons seen by LINC-NIRVANA scientific camera with K' filter inserted. The SX side was collimated, while the DX still needed fine-tuning. *Right*: 5th magnitude star observed on the HWS Patrol cameras (on-axis).

The main goal of the LINC-NIRVANA's commissioning nights was, needless to say, to test MFoV MCAO on multiple stars on sky. As one can expect, the commissioning of the 4 WFSs is not proceeding at the same speed. However, so far, we have been able to achieve the basic goals on all WFSs, such as multiple NGSs acquisition and auto-guiding for minutes to hours.

The first arm to be commissioned was SX. The strategy involved commissioning the GWS SX first and the HWS as a second task, since the tip-tilt signal would exceed the dynamic range of the Xinetics DM, unless the ground-layer correction is performed.

On March 29th 2017, looking at Field 08.9 RA +11.9 Dec, with 1.0" seeing and 1.5 airmasses, we were able to close the ground-layer loop on the SX arm, correcting up to 20 modes on 5 stars and record an image on the scientific camera in the K' filter (Figure 5 right). The chosen asterism featured 5 stars well distributed on the GWS FoV around the scientific field, as shown in Figure 5 left. One NGSs was about 7th magnitude (R-band) while the other four were of about 10.5-11th magnitude (R-band). The FWHM of the SX side star after correction is 0.47", meaning a correction of almost a factor 2 with respect to the uncorrected image (0.93"), consistent with the literature [17].

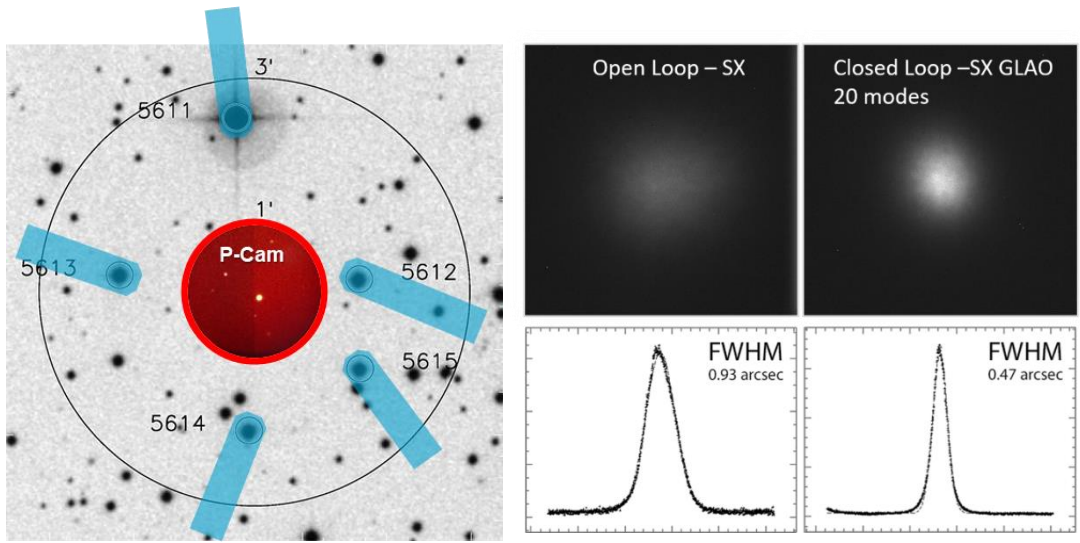

Figure 5: *Left:* stellar field used to perform GLAO on 5 NGSs using the SX side. The 3' radius circle indicates the GWS annular FoV, while the inner 1' radius the HWS one. Blue items represent the SEs probes. The central image shows the real stellar field, as seen through the Patrol camera after proper rotation of the image. *Left:* K' scientific camera. The left side shows the open loop PSF, while the right side shows the PSF obtained with the 5 star GLAO correction. The FWHM shows a correction of about a factor 2, with respect to the seeing-limited condition.

On June 11th 2017 (latest LN on-sky date at the time of this writing), looking at Field 19.6 RA +35.7 Dec, with 1.3'' seeing, we performed the first effective attempt of MFoV MCAO on two pairs of stars (one for each sensor), with the asterism shown in Figure 6. The GWS star magnitude were about 9th and 11th (R-band), while the HWS ones were about 8th magnitude. The modest improvement on GLAO image (FWHM ~ 0.74" vs 0.94" in open-loop) is probably due to the fact that only 2 NGSs were used, although we corrected up to 40 modes. Concerning the FWHM obtained in closed-loop with 10 modes on HWS, it was a first attempt, where the seeing was quite high, the differential focus of the camera optimization was not performed and NCPA were not taken into account.

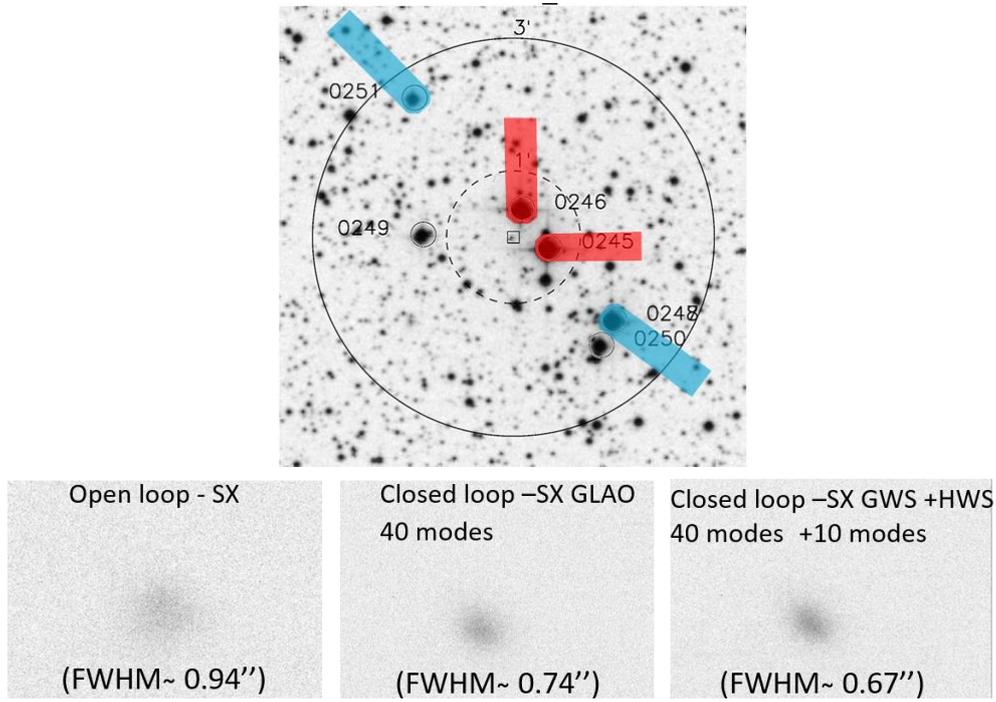

Figure 6: *Top:* stellar field used to perform MFoV MCAO on 2 + 2 NGSs using the SX side. The 3' radius circle indicates the GWS annular FoV, while the inner 1' radius the HWS one. Blue items represent the GWS SEs probes, while red ones the HWS probes. The central square shows the scientific FoV. *Bottom*: On the left side is shown the open loop PSF (0.94" FWHM), in the center the closed loop obtained with just GWS correction 2 stars (0.74") and on the right side the correction also for HWS (0.67")

## 5. LESSONS LEARNED AND CONCLUSIONS

The LINC-NIRVANA's size and weight, added to its very high complexity in terms of the number of components and needed synergy between all of them, can be easily taken as a realistic example of what can be expected for the alignment and commissioning of the upcoming ELT AO instruments.

Concerning its alignment, it has to be stressed that the telescope mirrors were aligned to the instrument and not vice-versa. This has to be considered both in the design of the future ELTs and of the instruments. At the same time, as obvious as it might sound, AIV planning, including alignment to the telescope, needs to be developed.

Furthermore, MCAO will be a key aspect for science in the next generation ELT instruments. Therefore, LN's commissioning phase can teach some lessons, from the obvious need to perform day-time calibration, to the development of algorithms to improve time efficiency and reduce the use of telescope time, such as the ones developed for LN commissioning, detailed in other papers presented at this conference [14][15][16]. Although the Pathfinder experience allowed us to save some time during LN commissioning, mainly in the GWS calibration procedure and in the interaction with the telescope, the main lesson learned performing these activities is the fact that the commissioning of an AO system made of 4 WFSs with 40 pyramids confirmed to be as challenging as we foresaw it.

Nevertheless, in about 3.5 nights of commissioning, spreaded over 7 days, between March and June 2017, we were able to prove GLAO correction with almost a factor 2 gain on the scientific image, closing the loop on the GWS SX using 5 NGSs and to attempt the first MFoV MCAO on 2 pairs of stars, closing the loop correcting up to 40 modes on GWS SX and up to 10 modes on HWS SX. In the latter case we obtained a modest correction, due to the low number of stars used and the lack of time for optimization of parameters.

In the next run, which is scheduled for the end of October 2017, MFoV MCAO and scientific image improvement with a larger number of NGSs is the goal.

## AKNOWLEDGMENTS

We wish to acknowledge the INAF-Arcetri team for all Adaptive Secondary Mirror related assistance, J. Hill for its active participation to alignment and observations activities and the LBTO staff for helping us in many different ways.